\documentclass[preprint,superscriptaddress,showpacs,tightenlines]{revtex4}
\usepackage{amssymb}
\usepackage{amsmath}
\usepackage{graphicx,bm}

\input{tcilatex}

\begin{document}

\title{Magneto-orientation and quantum size effect in SP-STM conductance in
the presence of a subsurface magnetic cluster }
\author{Ye.S. Avotina}
\affiliation{B.I. Verkin Institute for Low Temperature Physics and Engineering, National
Academy of Sciences of Ukraine, 47, Lenin Ave., 61103, Kharkov,Ukraine.}
\author{Yu.A. Kolesnichenko}
\affiliation{B.I. Verkin Institute for Low Temperature Physics and Engineering, National
Academy of Sciences of Ukraine, 47, Lenin Ave., 61103, Kharkov,Ukraine.}
\author{J.M. van Ruitenbeek}
\affiliation{Kamerlingh Onnes Laboratorium, Universiteit Leiden, Postbus 9504, 2300
Leiden, The Netherlands.}

\begin{abstract}
The influence of a single magnetic cluster in a non-magnetic host
metal on the spin current $\mathbf{j}^{(s) }$ and the charge
current $\mathbf{j}$ in the vicinity of a ferromagnetic STM tip is
studied theoretically. Spin-flip processes due to electron
interaction with the cluster are taken into account. We show that
quantum interference between the partial waves injected from the
STM tip and those scattered by the cluster results in the
appearance of components perpendicular to the initial polarization
of the spin current $\mathbf{j}^{( s) }$, which obtain a strongly
inhomogeneous spatial distribution. This interference produces
oscillations of the conductance as a function of the distance
between the contact and the cluster center. The oscillation
amplitude depends on the current polarization. We predict a strong
magneto-orientational effect: the conductance oscillations may
grow, shrink, or even vanish for rotation of the cluster magnetic
moment $\mathbf{\mu }_{\mathrm{eff}}$ by an external magnetic
field. These results can be used for the determination of the $
\mathbf{\mu }_{\mathrm{eff}}$ for magnetic clusters below a metal
surface.
\end{abstract}

\maketitle

\section{Introduction}

Subsurface defects, such as impurities and vacancies, result in
oscillations of the conductance as a function of the position of a
Scanning Tunnelling Microscope (STM) tip relative to the defect
position (see, for example, \cite{Quass,babble,Kurnosikov}). These
Friedel-like oscillations originate from interference between
electron waves directly transmitted through the contact and waves
that are scattered by the defect and reflected back by the
contact. The theory of STM conductance in the presence of a single
non-magnetic point-like defect below a metal surface in the
vicinity of the contact has been developed in
Refs.~\cite{Kob,Avotina1}. First the signature of Fermi-surface
anisotropy in STM conductance in the presence of defects had been
analyzed theoretically in detail in Ref.\cite{Avotina2}. In the
paper \cite{Avotina2a} the general results of Ref.\cite{Avotina2}
was applied for the theoretical investigation of the conductance
of a tunnel point contact of noble metals in the presence of a
single defect below surface. A pattern of the conductance
oscillations, which can be observed by the method of scanning
tunneling microscopy, was obtained for different orientations of
the surface for the noble metals \cite{Avotina2a}. Recently it had
been confirmed experimentally that Fermi surfaces can be imaged in
real space with a low-temperature scanning tunneling microscope
when subsurface point scatterers are present \cite{Waismann}. The
effect of Kondo scattering by a subsurface magnetic impurity on
the conductance of a tunnel point contact has been analyzed
theoretically in Ref.~\cite{Avotina4}.

The applicability of STM can be extended for the study of magnetic objects
below the surface of a conductor when a magnetic material is used for the
STM tip such that the electric current is spin polarized (SP) (for review of
SP-STM see \cite{Bode}). An important feature of SP-STM is that the spin
polarized current influences a magnetic object in a non-magnetic matrix,
producing so-called spin transfer torque (for review, see \cite{Ralph}). For
example, near a point contact, where the current density is large, the spin
transfer torque can be strong enough to reorient the magnetization of
ferromagnetic layers in magnetic multilayers \cite{Tsoi}. Such
investigations are very important for the development of innovative
high-density data storage technologies.

In this paper we consider theoretically the conductance of a tunnel
point-contact between magnetic and non-magnetic metals in a SP-STM geometry.
A magnetic cluster is embedded in the non-magnetic metal in the vicinity of
the contact. The changes in the spin-polarized current and the spin transfer
torque that influences the magnetic moment of the cluster are analyzed. We
study the dependence of the amplitude of the conductance oscillations as a
function of the STM tip position, on the relative orientation between the
spin polarization of the tunnelling electrons and the magnetic moment of the
cluster $\mathbf{\mu }_{\mathrm{eff}}$.

\section{Model of the contact and basic equations}

Let us consider a tunnel contact between a semi-infinite
half-space $z\geq 0$ of a nonmagnetic metal (the sample) and a
sharp tip of a magnetic conductor (Fig.~1). A voltage $V$ is
applied between the tip and the sample. The electrical (and spin)
current flows through a small region of the surface at $z=0$ near
the tip apex that is closest to the sample. This system models the
geometry of a SP-STM experiment. The tip magnetization (in real
SP-STM the magnetization of the last atom \cite{Bode}), which we
choose along the $ z $-direction, defines the direction of the
polarization of the tunnel current. Such magnetization can be
obtained, for example, for a Fe/Gd-coated W STM tip \cite{Kub}. In
the non-magnetic metal we place a spherical single-domain magnetic
cluster having a magnetic moment $\mu _{\mathrm{eff}}$ (Fig.~1).
As first predicted by Frenkel and Dorfman \cite{Frenkel} particles
of a ferromagnetic material are expected to organize in a single
magnetic domain below a critical particle size (a typical value
for this critical size for Co is about 35nm). Depending on the
size and the material, the magnetic moments of such particles can
be $\mu _{\mathrm{eff}}\sim 10^{2}-10^{5}\mu _{\mathrm{B}}$, where
$\mu _{\mathrm{B}}$ is the Bohr magneton \cite{Chikazumi}. Below,
we only consider electron scattering by the magnetic cluster,
assuming the mean free paths for all other processes (electron
spin-diffusion length, electron-phonon mean free path and others)
are much larger than the distance between the contact and the
cluster center $r_{0}$.

\begin{figure}[tbp]
\includegraphics[width=10cm,angle=0]{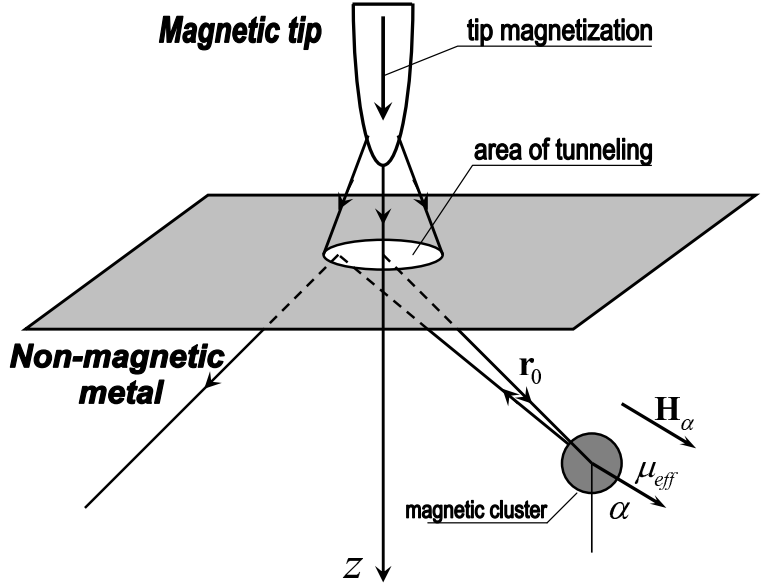}
\caption{Model of the contact in a SP-STM configuration, with a subsurface
magnetic cluster near the tunnel contact point. Spin-polarized electrons
tunnel into the non-magnetic metal in the small area below the STM tip. The
trajectories of electrons that are scattered by the spherical magnetic
cluster are shown schematically.}
\end{figure}

In order to describe the spin-polarized electron states of our
system we use the approach proposed in the works of Slonczewski
and Berger \cite {Slonczewski,Berger}: all calculations are
performed by means of independent single particle spinor wave
functions $\widehat{\Psi }(\mathbf{r}_{l};\sigma _{l})$ of
electrons with opposite spin directions, where $\mathbf{r}_{l}$
and $\sigma _{l}$ are the position vector and the spin direction,
respectively, for each spin orientation $l=1,2$ . We use the
representation $ \sigma _{1,2}=\uparrow ,\downarrow $, in which
the spin projection on the $z$ axis, $s_{z}=\pm \frac{1}{2},$ is
used. This approach corresponds to reducing the many-particle
problem of a partially polarized electron system with non-zero
average spin to a double-particle problem for electrons in a pure
(completely polarized) spin state. Neglect of electron-electron
interactions enables us to separate the double particle
Schr\"{o}dinger equation into two independent equations for
$\widehat{\Psi }(\mathbf{r} _{l};\sigma _{l}).$ In our case this
separation is valid if $\mu _{\mathrm{ eff}}\gg \mu
_{\mathrm{B}},$ i.e. the electron-electron exchange interaction is
negligible compared to electron exchange interaction with the
cluster. Generally, the moment $\mathbf{\mu }_{\mathrm{eff}}$ of
the cluster in a non-magnetic metal takes an arbitrary direction.
This direction (the angle $ \alpha $ in Fig.~1) can be held fixed
by an external magnetic field $\mathbf{ H}_{\alpha }$, the value
of which is estimated as $H_{\alpha }\simeq T/\mu _{
\mathrm{eff}}$, where $T$ is the temperature (see, for example,
Ref.~\cite {Vonsovsky}). For $\mu _{\mathrm{eff}}\simeq 10^{2}\mu
_{\mathrm{B}}$ and $ T\sim 1$K the field $H_{\alpha }$ is of the
order of $0.1$T. We assume that $ H_{\alpha }$ is much smaller
than the magnetocrystalline anisotropy field of the STM tip, i.e.
the direction of the external magnetic field controls the
direction of the cluster magnetic moment but its influence on the
spin-polarization of the STM current is negligible. If the
magnetic moment $ \mathbf{\mu }_{\mathrm{eff}}$ of the cluster is
'frozen' by the field $ H_{\alpha }$ the problem becomes a
stationary one. Also we take the distance between the contact and
the cluster $r_{0}$ to be much smaller than the radius
$r_{\mathrm{H}}=cp_{\mathrm{F}}/eH_{\alpha }$ of the electron
trajectory in the applied external magnetic field $H_{\alpha }$,
and we do not take into account trajectory magnetic effects, which
have been analyzed in Ref.~\cite{Avotina3}. The condition
$r_{0}\ll r_{\mathrm{H}}$ together with inequality $\mu
_{\mathrm{B}}H_{\alpha }/\varepsilon \ll 1$ ($ \varepsilon $ is
the electron energy) allow neglecting the magnetic field in the
single-electron Hamiltonian and considering $\mathbf{H}_{\alpha }$
as an independent external parameter.

The external magnetic field may result in a modulation of the
tunnel current due to electron spin precession \cite{Jedema}. For
our problem such precession would become noticeable when the
transit time for the electron motion from the contact to the
cluster $t\simeq r_{0}/v_{\mathrm{F}}$ is larger than $1/\omega
_{\mathrm{L}},$ where $v_{\mathrm{F}}$ and $\omega _{ \mathrm{L}}$
are the Fermi velocity and Larmor frequency. The inequality
mentioned above, $r_{0}\ll r_{\mathrm{H}}$, is equivalent to the
condition $ \omega _{\mathrm{L} }t\ll 1,$ so that the effect of
electron spin precession is negligible.

Thus, under the assumptions outlined above the spinor wave
functions $ \widehat{\Psi }(\mathbf{r}_{l};\sigma _{l})$ satisfy
the one-electron Schr \"{o}dinger equation
\begin{equation}
\left( -\frac{\hbar ^{2}}{2m^{\ast }}\nabla _{l}^{2}-\varepsilon
\right) \widehat{I}\widehat{\Psi }(\mathbf{r}_{l};\sigma
_{l})=-\widehat{U}(\mathbf{r }_{l})\widehat{\Psi
}(\mathbf{r}_{l};\sigma _{l}),  \label{Shrod}
\end{equation}
where $m^{\ast }$ is the effective mass of the electrons, and
$\widehat{U}( \mathbf{r})$ is the interaction potential of the
electrons with the cluster. The matrix $\widehat{U}$ \ consists of
two parts, describing the potential as well as the magnetic
scattering,
\begin{equation}
\widehat{U}(\mathbf{r})=\left( g\widehat{I}+\frac{1}{2\mu
_{\mathrm{B}}}J \mathbf{\ \mu
}_{\mathrm{eff}}\widehat{\mathbf{\sigma }}\right) D(\left\vert
\mathbf{r-r}_{0}\right\vert ),  \label{V}
\end{equation}
where $g$ is the constant describing the non-magnetic part of the
interaction (for $g>0$ the potential is repulsive), $J$ is the
constant of exchange interaction, $\mathbf{\mu
}_{\mathrm{eff}}=\mathbf{\mu }_{\mathrm{ eff}}(\sin \alpha ,0,\cos
\alpha )$ is the magnetic moment of the cluster, $
\widehat{\mathbf{\sigma }}=\left( \widehat{\sigma
}_{x},\widehat{\sigma } _{y},\widehat{\sigma }_{z}\right) $ with
$\widehat{\sigma }_{\mu }$ the Pauli matrixes, and $\widehat{I}$
is the unit matrix. $D(\left\vert \mathbf{ r-r}_{0}\right\vert )$
is a spherically symmetric function localized within a region of
characteristic radius $r_{\mathrm{D}}$ centered at the point $
\mathbf{r}=\mathbf{r}_{0}$, which satisfies the normalization
condition
\begin{equation}
\int d\mathbf{r}^{\prime }D(\mathbf{r}^{\prime })=1.
\end{equation}
Equation (\ref{Shrod}) must be supplemented with the common boundary
conditions for continuity of the wave function and its normal derivative on
the metal surface.

We assume that the potential $\widehat{U}( \mathbf{r}) $ is small
and use perturbation theory. As a first step we should find the
solutions $\widehat{ \Psi }^{( 0) }( \mathbf{r}_{l};\sigma _{l}) $
of Eq. (\ref{Shrod}) for $ \widehat{U}( \mathbf{r}) =0.$
Generally, this solution depends on the model chosen to represent
the tunnel barrier. For any suitable model for the potential
barrier the wave functions $\widehat{\Psi } ^{( 0) }( \mathbf{r}
_{l};\sigma _{l}) $ describe the spreading of the electron waves
in the metal from the small contact region on the surface. Here,
we use the results of Refs.~\cite{Avotina4,KMO}, in which the
tunnel contact is modelled in the form of a circular orifice of
radius $a$, with a large amplitude potential barrier $U_{0}\delta
( z) $.

In order to describe the spin polarization of the STM current we introduce
different magnitudes for the wave vector $\mathbf{k}_{\sigma }$ for spin-up
and spin-down electrons (for the same energy $\varepsilon $) before
tunnelling from the tip. This difference results in different amplitudes $t(
\mathbf{k}_{\sigma }) $ of the electron waves injected into the non-magnetic
metal for different directions of the spin \cite{Avotina4,KMO},
\begin{equation}
t_{\sigma }( \mathbf{k}_{\sigma }) \approx \frac{\hbar ^{2}k_{\sigma }\cos
\vartheta}{im^{\ast }U_{0}};\quad |t_{\sigma }|\ll 1.  \label{t}
\end{equation}
Here, $\vartheta $ is the angle between wave vector $\mathbf{k}_{\sigma }$
and the normal to the surface $z=0$, pointing into the sample. The total
effective polarization $P_{\mathrm{eff}} $ of the current depends on the
difference between the probabilities of tunnelling for different $\sigma$,
\begin{equation}
P_{\mathrm{eff}}( \varepsilon ) =\frac{\left\vert t_{\uparrow
}\right\vert ^{2}-\left\vert t_{\downarrow }\right\vert
^{2}}{\left\vert t_{\uparrow }\right\vert ^{2}+\left\vert
t_{\downarrow }\right\vert ^{2}}=\frac{ k_{\uparrow
}^{2}-k_{\downarrow }^{2}}{k_{\uparrow }^{2}+k_{\downarrow }^{2}}
.  \label{P}
\end{equation}
The functions $\widehat{\Psi }^{( 0) }( \mathbf{r}_{l};\sigma _{l}) $ for
each spin direction have the same spatial dependence as those for a contact
between non-magnetic metals. In following calculations we use the asymptotic
expression for $\widehat{\Psi }^{( 0) }( \mathbf{r}_{l};\sigma _{l}) $ valid
for $ka\ll 1$ \cite{Avotina4}
\begin{equation}
\widehat{\Psi }^{( 0) }( \mathbf{r};\sigma ) =t_{\sigma }\Psi ^{( 0) }(
\mathbf{r}) \widehat{\psi } _{\sigma },  \label{Psi0}
\end{equation}
where
\begin{equation}
\Psi ^{( 0) }( \mathbf{r}) =\frac{\left( ka\right)
^{2}z}{2ikr^{2}} e^{ikr}\left( 1+\frac{1}{ikr}\right),
\label{psi00}
\end{equation}
$\widehat{\psi }_{\sigma }$ is the spinor
\begin{equation}
\widehat{\psi }_{\uparrow }=\left(
\begin{array}{c}
1 \\
0
\end{array}
\right), \quad \widehat{\psi }_{\downarrow }=\left(
\begin{array}{c}
0 \\
1
\end{array}
\right),
\end{equation}
and $k=\sqrt{2m^{\ast }\varepsilon }/\hbar $ is the magnitude of
the electron wave vector $\mathbf{k}$ in the non-magnetic metal.
Note that the wave function $\Psi ^{( 0) }( \mathbf{r}) $
(\ref{psi00}) is zero in all points on the surface $z=0$, except
the point $r=0$ (at the contact) where it diverges. This
divergence is the result of taking the limit $a\rightarrow 0$ in
the integral expressions for $\Psi ^{( 0) }( \mathbf{r}) $ \cite
{Avotina1,KMO}. Yet, Eq.~(\ref{psi00}) gives a finite value for
the total charge current through the contact as obtained over a
half-sphere of radius $ r$ with its center in the point
$\mathbf{r}=0$ for $r\rightarrow 0$.

The solution to Eq.~(\ref{Shrod}) in linear approximation in the
potential $ \widehat{U}$, Eq.~(\ref{V}), can be written as
\begin{equation}
\widehat{\Psi }(\mathbf{r};\sigma )=t_{\sigma }\left[ \Psi
^{(0)}(\mathbf{r}) \widehat{\psi }_{\sigma }+\left( \left(
\widetilde{g}\pm \widetilde{J}\cos \alpha \right) \widehat{\psi
}_{\sigma }+\widetilde{J}\sin \alpha \widehat{ \psi }_{-\sigma
}\right) \Psi ^{(1)}(\mathbf{r})\right] ;  \label{wf}
\end{equation}
where the sign $\left( \pm \right) $ corresponds to $\sigma =\uparrow
,\downarrow $. We have introduced the notation
\begin{equation}
\widetilde{g}=\frac{2m^{\ast }k}{\hbar ^{2}}g,\quad
\widetilde{J}=\frac{ m^{\ast }k}{\mu _{\mathrm{B}}\hbar ^{2}}J\mu
_{\mathrm{eff}}  \label{gJ}
\end{equation}
for the dimensionless constants of interaction for potential and magnetic
scattering, respectively. The wave function scattered by the cluster is
given by
\begin{equation}
\Psi ^{(1)}(\mathbf{r})=-\frac{1}{k}\int d\mathbf{r}^{\prime
}G_{0}^{+}( \mathbf{r,r}^{\prime })D(\left\vert \mathbf{r}^{\prime
}\mathbf{-r} _{0}\right\vert )\Psi ^{(0)}(\mathbf{r}^{\prime }),
\label{psi1}
\end{equation}
which undergoes specular reflection at the metal surface. Aiming
for a solution for the wave function $\widehat{\Psi }_{\mathbf{\
\sigma }}(\mathbf{ r})$ in first approximation in the small
parameter $|t_{\sigma }|\ll 1$ we substitute the electron Green
function $G_{0}^{+}(\mathbf{r,r}^{\prime })$ of the non-magnetic
half-space in Eq.~(\ref{wf}),
\begin{equation}
G_{0}^{+}(\mathbf{r,r}^{\prime })=G_{00}^{+}(\left\vert
\mathbf{r}-\mathbf{r} ^{\prime }\right\vert
)-G_{00}^{+}(\left\vert \mathbf{r}-\widetilde{\mathbf{r }}^{\prime
}\right\vert ),  \label{G+0}
\end{equation}
where $\widetilde{\mathbf{r}}=\left( \mathbf{\rho },-z\right) $ is
the mirror image of the point $\mathbf{r}$ relative to the metal
surface, and $ G_{00}^{+}$ is the Green function for free
electrons,
\begin{equation}
G_{00}^{+}(R)=-\frac{e^{ikR}}{4\pi R};\quad R=\left\vert
\mathbf{r}-\mathbf{r }^{\prime }\right\vert .  \label{G00}
\end{equation}

The wave function (\ref{wf}) enables calculation of the charge current
density $\mathbf{j}$\ and spin current density $\mathbf{j}^{( \mu ) }$, and
the expectation value of the spin $\mathbf{s.}$ They are obtained as the
sums of independent contributions of the two electron groups $\left(
l=1,2\right) $
\begin{equation}
\mathbf{j}( \mathbf{r}) =\frac{e\hbar }{m^{\ast
}}\sum\limits_{l=1,2}\func{Im }\left( \widehat{\Psi }\nabla
\widehat{\Psi }^{+}\right) _{\mathbf{r}_{1}=
\mathbf{r}_{2}=\mathbf{r}};  \label{j}
\end{equation}
\begin{equation}
\mathbf{j}^{( \mu ) }( \mathbf{r}) =\frac{i\hbar }{2m^{\ast }}
\sum\limits_{l=1,2}\left[ \left( \nabla \widehat{\Psi }^{+}\right)
\sigma _{\mu }\widehat{\Psi }-\widehat{\Psi }^{+}\sigma _{\mu
}\left( \nabla \widehat{\Psi }\right) \right]
_{\mathbf{r}_{1}=\mathbf{r}_{2}=\mathbf{r}}; \label{js}
\end{equation}
\begin{equation}
\mathbf{s}( \mathbf{r}) =\sum\limits_{l=1,2}\left( \widehat{\Psi
}^{+} \widehat{\mathbf{\sigma }}\widehat{\Psi }\right)
_{\mathbf{r}_{1}=\mathbf{r} _{2}=\mathbf{r}}.  \label{s}
\end{equation}
The Eqs.~(\ref{j})-(\ref{s}) with wave function (\ref{wf}) describe the
so-called tunnelling contributions (see, \cite{Slonczewski}). They are
proportional to the tunnelling probability and define the measurable
quantities which can be obtained after averaging over the energies of the
tunnelling electrons and wave vector directions (see, Sec.~III).

In absence of the cluster $s_{x}=s_{y}= 0,$ and the local
magnetization $ \mathbf{s}_{0}( \mathbf{r}) $ due to itinerant
spin polarized electrons is oriented along the $z$-axis. The spin
polarization in zeroth approximation, $ s_{z0}( \mathbf{r}) ,$
which is calculated from the wave function (\ref{Psi0} ), is a
monotonic function of coordinates
\begin{equation}
s_{z0}( \mathbf{r}) =\left( \left\vert t_{\uparrow }\right\vert
^{2}-\left\vert t_{\downarrow }\right\vert ^{2}\right) \left(
\frac{kza^{2}}{ 2r^{2}}\right) ^{2}\left( 1+\frac{1}{\left(
kr\right) ^{2}}\right) . \label{sz0}
\end{equation}

The electron scattering by the spin-depended potential (\ref{V})
changes the value and the direction of the vector $\mathbf{s}_{0}(
\mathbf{r}).$ Components $s_{x}$ and $s_{y}$ appear due to
electron scattering by the cluster and they are subject to quantum
interference between transmitted and scattered waves. As a result
of the interference the spin components perpendicular to the
$z$-axis are oscillatory functions of the coordinates, while
$s_{z0}$ acquires a small oscillatory component proportional to $
\widetilde{J}^{2}$.
\begin{equation}
\left\{
\begin{array}{c}
s_{x}( \mathbf{r}) \\
s_{y}( \mathbf{r})
\end{array}
\right\} =\left( \left\vert t_{\uparrow }\right\vert ^{2}\pm \left\vert
t_{\downarrow }\right\vert ^{2}\right) \widetilde{J}\sin \alpha \left\vert
\Psi ^{( 0) }( \mathbf{r}) \right\vert \left\vert \Psi ^{( 1) }( \mathbf{r})
\right\vert \left\{
\begin{array}{c}
\cos \\
\sin
\end{array}
\right\} \left( \varphi _{1}( \mathbf{r}) -\varphi _{0}( \mathbf{r}) \right)
\end{equation}
Here $\left\vert \Psi ^{( i) }( \mathbf{r}) \right\vert $ and $\varphi _{i}(
\mathbf{r}) $ are the absolute values and the phases of the coordinate part
of the wave functions (\ref{wf}) (see, Eqs.~(\ref{psi00}) and (\ref{psi1})).

Figure~2 shows the spacial distribution of the $x$-component of the
normalized spin density $s_{x}( \mathbf{r}) /c_{0}$ in the vicinity of the
contact. The normalization constant $c_{0}=\left( \left\vert t_{\uparrow
}\right\vert ^{2}+\left\vert t_{\downarrow }\right\vert ^{2}\right)
\widetilde{J}\left( ka\right) ^{4}\sin \alpha /16\pi $. The figure
demonstrates that the spin component $s_{x}( \mathbf{r}) $ changes sign in
the space of the normal metal. The sign of $s_{x}$ depends on the difference
of the phases $\varphi _{i}( \mathbf{r}) $.
\begin{figure}[tbp]
\includegraphics[width=10cm,angle=0]{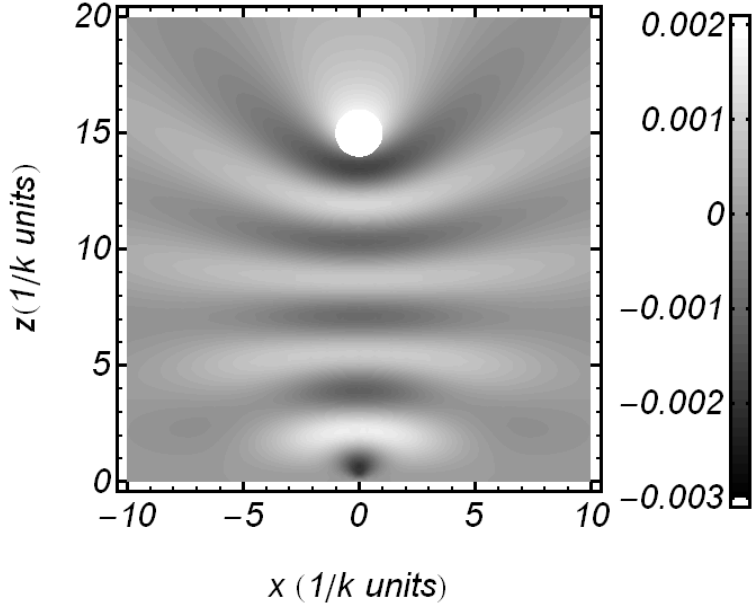}
\caption{Gray-scale plot of the spacial distribution of the
$x$-component of the spin density, $s_{x}( \mathbf{r}) /c_{0}.$
The coordinate plane $xz $ in real space crosses the contact and
the cluster of the radius $r_{\mathrm{D} }=k^{-1}$, with its
center in the point $\mathbf{r}_{0}=\left( 0,0,15\right) k^{-1}.$}
\end{figure}
In Fig.~3 we present a vector plot of the $x$-component of the spin current
density $\mathbf{j}^{( x) }$. An intricate image of the distribution in
orientation of $\mathbf{j }^{( x) }$ is visible. Note the lines at which the
direction of the vector $\mathbf{j}^{( x) }$ is inverted.
\begin{figure}[tbp]
\includegraphics[width=10cm,angle=0]{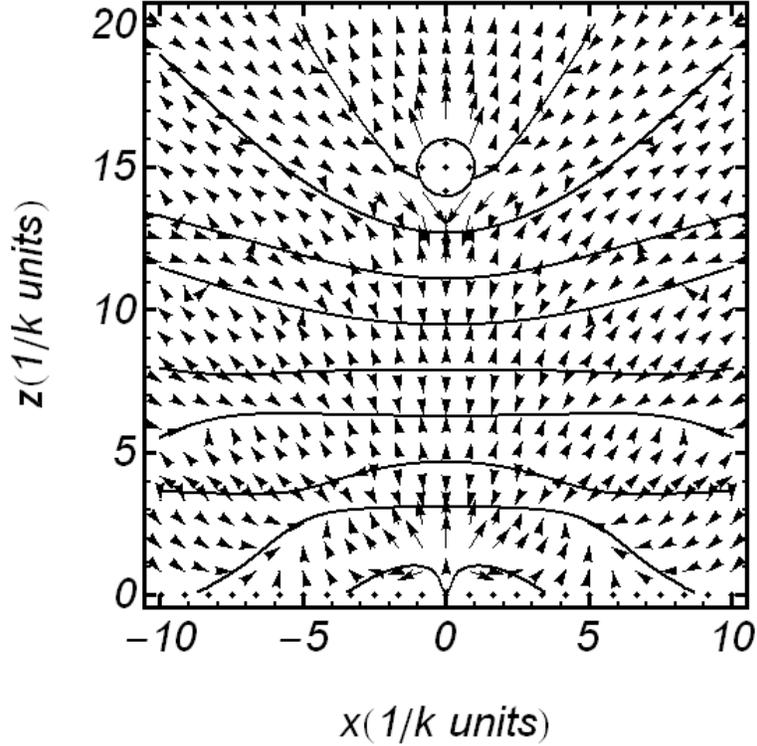}
\caption{Vector plot of the direction of the $x$-component of spin current
density, $\mathbf{j}^{( x) }$. The contour lines correspond to $j_{z}^{( x)
}=0.$ The plane $xz$ crosses the contact and the cluster; as in Fig.~2 we
have chosen $r_{\mathrm{D}}=k^{-1}$ and $\mathbf{r}_{0}=\left( 0,0,15\right)
k^{-1}.$}
\end{figure}

\section{Conductance of the contact, spin current, and spin-transfer torque}

\begin{figure}[tbp]
\includegraphics[width=10cm,angle=0]{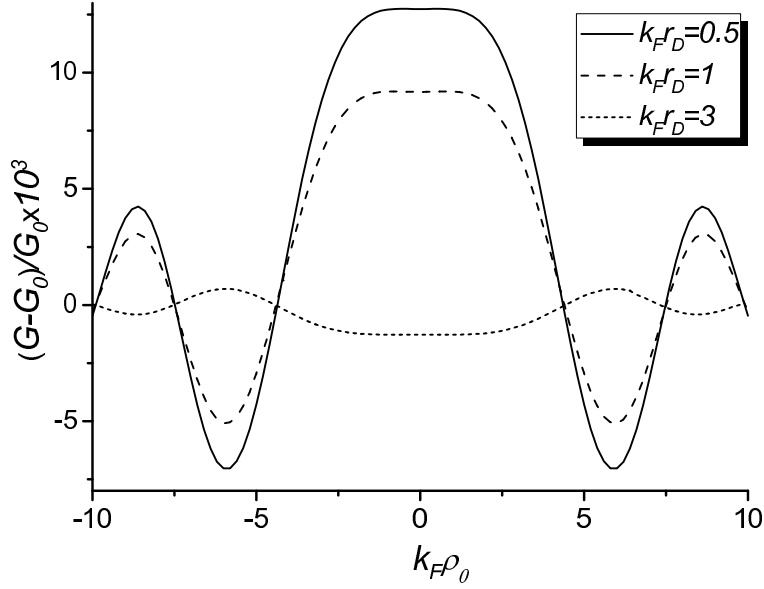}
\caption{Dependence of the the oscillatory part of the conductance
on the tip position on the metal surface for different cluster
diameters. The $ \protect\rho _{0}$-coordinate is measured from
the point $\protect\rho _{0}=0 $ at which the contact is situated
directly above the cluster; $ r_{0}=(0,0,10)/k_{F};$
$\widetilde{g}=0.5;$ $\widetilde{J}=2.5;$ $ P_{eff}=0.4 $;
$\protect\alpha =0$. } \label{Fig4}
\end{figure}

The total current through the contact can be evaluated by
integration of the charge current density
$\mathbf{j}(\mathbf{r})$, Eq.~(\ref{j}), over a half-sphere
centered at the point contact $r=0$ and covering the contact at $
z>0$, and integration over all directions of the electron wave
vector on the Fermi surface $\varepsilon =\varepsilon _{F}$ . In
the Ohm's law approximation and at zero temperature the total
current through the contact $ I$ can be written as
\begin{equation}
I=eV\rho (\varepsilon _{F})r^{2}\int\limits_{\varepsilon
=\varepsilon _{ \mathrm{F}}}\frac{d\Omega _{\mathbf{k}}}{4\pi
}\int d\Omega \Theta (z)\Theta (k_{z})j_{r}(\mathbf{r}),
\label{Gint}
\end{equation}
where $d\Omega $ and $d\Omega _{\mathbf{k}}$ are elements of solid
angle in real and momentum space, respectively, $\rho (\varepsilon
_{F})$ is the electron density of states at the Fermi energy,
$\varepsilon _{\mathrm{F}}$, for one spin direction, $k_{z}$ is
$z$-component of the wave vector, $j_{r}( \mathbf{r})$ is the
radial component of $\mathbf{j}(\mathbf{r})$ (\ref{j}), and
$\Theta (x)$ is the Heaviside unit step function. After
integration of Eq.~(\ref{Gint}) the conductance $G$ of the contact
takes the form
\begin{equation}
G=\frac{I}{V}=G_{0}\left[ 1+\frac{6}{\pi }\left(
\widetilde{g}+P_{\mathrm{eff }}\widetilde{J}\cos \alpha \right)
W(\mathbf{r}_{0})\right] _{\varepsilon =\varepsilon _{\mathrm{F}}}
\label{G}
\end{equation}
where $G_{0}$ is the conductance of the contact in absence of the
cluster
\begin{equation}
G_{0}=\left( k_{\mathrm{F}\uparrow }^{2}+k_{\mathrm{F}\downarrow
}^{2}\right) \frac{e^{2}\hbar ^{3}\left( k_{\mathrm{F}}a\right) ^{4}}{72\pi
\left( m^{\ast }U_{0}\right) ^{2}}.  \label{G_0}
\end{equation}
Here, $k_{\mathrm{F}\sigma }$ and $k_{\mathrm{F}}$ are the absolute values
of electron wave vectors at the Fermi level in magnetic and non-magnetic
metals, respectively; $P_{\mathrm{eff}}$ is the effective spin polarization
of the current injected through the contact (see Eq.~(\ref{P})); the
constants $\widetilde{g}$ and $\widetilde{J}$ are given by Eqs.~(\ref{gJ}),
and
\begin{equation}
W(\mathbf{r}_{0})=\int d\mathbf{r}^{\prime }D(\left\vert
\mathbf{r}^{\prime } \mathbf{-r}_{0}\right\vert )\left(
\frac{z^{\prime }}{r^{\prime }}\right) ^{2}y_{1}(kr^{\prime
})j_{1}(kr^{\prime }),  \label{W}
\end{equation}
where $j_{l}(x)$ and $y_{l}(x)$ are the spherical Bessel
functions. Equation~(\ref{G}) coincides with the result for a
tunnel point contact between non-magnetic metals \cite{Avotina4}
when $P_{\mathrm{eff}}=0$ and $ kr_{\mathrm{D}}\ll 1$. When the
radius of action $r_{\mathrm{D}}$ of the function $D(\left\vert
\mathbf{r-r}_{0}\right\vert )$ is much smaller than the distance
between the contact and the cluster center $r_{0},$ $W(\mathbf{r
}_{0})$ is an oscillatory function of $kr_{0}$ for
$kr_{\mathrm{D}}\geq 1,$ but the oscillation amplitude is reduced
as a result of superposition of waves scattered by different
points of the cluster. The integral $W(\mathbf{r }_{0})$,
(\ref{W}) can be calculated asymptotically for $r_{0}\gg
r_{\mathrm{ D}},$ $kr_{0}\gg 1,$ and $kr_{\mathrm{D}}\gtrsim 1.$
For a homogeneous spherical potential $D(\left\vert
\mathbf{r}\right\vert )=V_{D}^{-1}\Theta (r_{\mathrm{D}}-r)$
($V_{D}$ is the cluster volume) the function $W(\mathbf{r }_{0})$
takes the form
\begin{equation}
W(\mathbf{r}_{0})\simeq \frac{3}{2}\left(
\frac{z_{0}}{r_{0}}\right) ^{2} \frac{\sin 2kr_{0}}{\left(
2kr_{0}\right) ^{2}}\frac{j_{1}(kd)}{kd}, \label{Wasym}
\end{equation}
where $d=2r_{\mathrm{D}}$ is the cluster diameter. The last factor
in Eq.~( \ref{Wasym}) describes the quantum size effect related
with electron reflections by the cluster's boundary. Such
oscillations may exist, if the cluster boundary is sharp on the
scale of the electron wave length. Fig.~\ref {Fig4} shows the
dependence of the amplitude of the conductance oscillations on the
cluster diameter. It demonstrates that a $\pi$-phase shift may
occur resulting from interference of electron waves over a
distance of the cluster diameter.

In Eq.~(\ref{G}) the term proportional to $P_{\mathrm{eff}}$ takes
into account the difference in the probabilities of scattering of
electrons with different $\sigma $ by the localized magnetic
moment $\mu _{\mathrm{eff}}.$ It depends on the angle $\alpha $
between the tip magnetization and $\mathbf{ \mu }_{\mathrm{eff}}$
as $\cos \alpha .$ The same dependence was first predicted for a
tunnel junction between ferromagnets for which the magnetization
vectors are misaligned by an angle $\alpha $ \cite{Slonczewski} ,
and this was observed in SP-STM experiments \cite{Bode}.

Similar to the electrical conductance (\ref{G}), the total spin current for
each spin component can be calculated
\begin{equation}
I^{( z) }=\frac{G_{0}V}{e}\left[ P_{\mathrm{eff}}+\frac{6}{\pi
}\left( P_{ \mathrm{eff}}\widetilde{g}+\widetilde{J}\cos \alpha
\right) W( \mathbf{r} _{0}) \right] _{\varepsilon =\varepsilon
_{\mathrm{F}}};  \label{Isz}
\end{equation}
\begin{equation}
I^{( x) }=\frac{6G_{0}V}{e\pi }\sin \alpha \left[ \widetilde{J}W(
\mathbf{r} _{0}) \right] _{\varepsilon =\varepsilon
_{\mathrm{F}}}.  \label{Isx}
\end{equation}
For our choice of the vector $\mathbf{\mu }_{\mathrm{eff}}$, $I^{(
y) }=0.$ The value of the $z$-component of the spin current $I^{(
z) }$ (\ref{Isz} ) is determined to a large extent by the
polarization $P_{\mathrm{eff}}$ (\ref {P}) of the initial current.
The oscillatory part of $I^{( z) }$ is modified by the addition of
a term due to spin-flip processes on the cluster. The spin current
component perpendicular to initial direction of polarization, $
I^{( x) }$, has only a term that oscillates with $r_{0}$, and
which disappears when the magnetic moment $\mathbf{\mu
}_{\mathrm{eff}}$ is aligned with the $z$ direction.

The spin-transfer torque $\mathbf{T}$ acting on the magnetic
moment $\mathbf{ \mu }_{\mathrm{eff}}$ is given by
\begin{equation}
\mathbf{T=-}\frac{J}{\hbar }\int d\mathbf{r}^{\prime }D(\left\vert
\mathbf{r} ^{\prime }\mathbf{-r}_{0}\right\vert )\mathbf{\mu
}_{\mathrm{eff}}\times \left\langle \mathbf{s(\mathbf{r}^{\prime
})}\right\rangle \mathbf{;} \label{T}
\end{equation}
where
\begin{equation}
\left\langle \mathbf{s(\mathbf{r})}\right\rangle =eV\rho
(\varepsilon _{F})\int\limits_{\varepsilon =\varepsilon
_{\mathrm{F}}}\frac{d\Omega _{ \mathbf{k}}}{4\pi
}\mathbf{s(\mathbf{r});}
\end{equation}
$\mathbf{s(\mathbf{r})}$ is defined by Eq.~(\ref{s}). This torque
is related with the spin polarized electron tunnel current. In
linear approximation in $ \widetilde{J}$ only the spin density
contribution $s_{z0}\mathbf{(\mathbf{r}) }$ (\ref{sz0}), without
interaction with the cluster, should be taken for the calculation
of the torque (\ref{T}). In this approach $T_{x}=T_{z}=0.$ For
large $r_{0}\gg r_{\mathrm{D}}$ \ we obtain
\begin{equation}
T_{y}=\sin \alpha \frac{3G_{0}V}{e\pi }\left[
P_{\mathrm{eff}}\widetilde{J} \left(
\frac{z_{0}}{kr_{0}^{2}}\right) ^{2}\left( 1+\frac{1}{\left(
kr_{0}\right) ^{2}}\right) \right] _{\varepsilon =\varepsilon
_{\mathrm{F}}} \label{Ty}
\end{equation}
The dependence of $T_{y}$ (\ref{Ty}) on the angle $\alpha $ agrees
with the dependence of the spin transfer torque in tunnel
junctions between two ferromagnets with different directions of
magnetization \cite {Slonczewski,Berger}.

In this paper we do not aim to investigate the dynamics of the
cluster magnetic moment. We only note that once the spin-polarized
current-induced torque pulls the magnetic moment away from
alignment with $H_{\alpha }$, the cluster moment will start
precessing around the field axis. The Larmor frequency is defined
by the magnetic field due to combining the external field
$H_{\alpha }$ and the effective field produced by the polarized
current $H_{eff}\simeq -J\left\langle
\mathbf{s(\mathbf{r}}_{0}\mathbf{)} \right\rangle /g_{c}\mu
_{\mathrm{B}}$ ($g_{c}$ is the cluster 'g-factor'). The precession
of the cluster magnetic moment gives rise to a time modulation of
the SP-STM current as for clusters on a sample surface \cite
{Manoharan,Durkan}. The study of non-stationary effects provides a
further means of obtaining information on the cluster and the spin
polarization of the current inside the sample.

\section{Discussion}

In summary, we have studied theoretically the current and spin
flows through a tunnel point contact between magnetic and
non-magnetic metals when the tunnel current is spin polarized in
the geometry of SP-STM, Fig.~1. Electron spin flip processes due
to a magnetic cluster situated in the non-magnetic metal have been
taken into account. These processes result in the appearance of
components of the spin current density $\mathbf{j}^{( s) }(
\mathbf{r}) $ perpendicular to the source direction (taken along
the contact axis $z$), and a finite expectation value of the spin
$\mathbf{s}( \mathbf{r}) $. We have analyzed the contribution of
tunnelling electrons to the spacial distribution of $\mathbf{j}^{(
x,y) }( \mathbf{r}) $ and $s_{x,y}( \mathbf{r} ) .$ It is found
that these are non-monotonic functions of the coordinates and
undergo strong spacial oscillations originating from quantum
interference between partial waves transmitted through the contact
and those scattered by the cluster (see, Figs.~2, and 3).
Specifically, between the contact and the cluster there are
neighboring regions in which $\mathbf{j} ^{( x) }( \mathbf{r}) $
flows in opposite directions (Fig.~3).

The oscillatory correction, $\Delta G$, to the conductance $G_{0}$
of the ballistic tunnel point contact strongly depends on the
magnetic scattering ( \ref{G}),
\begin{equation}
\Delta G\sim \left( \widetilde{g}+P_{\mathrm{eff}}\widetilde{J}\cos \alpha
\right) \sin 2kr_{0},\quad kr_{0}\gg 1.
\end{equation}
The effective spin polarization $P_{\mathrm{eff}}$ of the
tunnelling electrons, and the dimensionless constants of potential
scattering $ \widetilde{g}$ and exchange scattering
$\widetilde{J}$ are given by Eqs.~( \ref{P}) and (\ref{gJ}).
Generally, for a single magnetic defect, which has a magnetic
moment of the order of one Bohr magneton, $\mu _{\mathrm{B}},$ the
spin part of electron scattering gives only a small contribution
to the electrical resistance. For a magnetic cluster with $\mu
_{\mathrm{eff}}\gg \mu _{\mathrm{B}},$ the exchange energy can be
larger than the energy of spin-independent interaction
($\widetilde{J}>\widetilde{g},$ see Eq.~(\ref {gJ})). An
interesting phenomenon may be found when $P_{\mathrm{eff}}
\widetilde{J}\geq \widetilde{g}$. A change of the direction of the
vector $ \mathbf{\mu }_{\mathrm{eff}}$ (the angle $\alpha $)
controlled by an external magnetic field is predicted to lead to a
change in the amplitude of the oscillations, and for certain
directions the amplitude may vanish, $ G_{osc}=0$. This large
magneto-orientational effect provides a new mechanism for
obtaining information on magnetic particles buried below a metal
surface. Note that this effect is observable only for a
spin-polarized current: if $t_{\uparrow }=t_{\downarrow },$ in
linear approximation in $ \widetilde{J}$ the changes in the
scattering amplitude for spin-up and spin-down electrons balance
each other and the magneto-orientational effect is absent.

As a consequence of spin flips due to the interaction of the electrons with
the cluster the oscillatory part $\Delta I^{( z) }$ of the spin current in
the original $z$ direction obtains a correction which depends on the
exchange constant $\widetilde{J}$ and the orientation of the magnetic moment
(\ref{Isz})
\begin{equation}
\Delta I^{( z) }\sim \left( P_{\mathrm{eff}}\widetilde{g}+\widetilde{J}\cos
\alpha \right) \sin 2k_{\mathrm{F}}r_{0},\quad k_{\mathrm{F}}r_{0}\gg 1.
\label{Iz}
\end{equation}
A component of the spin current perpendicular to the $z$
direction, $ I_{x}^{( s) }$, is formed only by scattered waves and
as the result of quantum interference it becomes an oscillatory
function of the distance $ r_{0}$ (\ref{Isx})
\begin{equation}
I^{( x) }\sim \widetilde{J}\sin \alpha \sin
2k_{\mathrm{F}}r_{0},\quad k_{ \mathrm{F}}r_{0}\gg 1.  \label{Ix}
\end{equation}
The spin currents (\ref{Iz}) and (\ref{Ix}) appear even in the case of
non-polarized current through the contact and they are related to the
magnetic scattering by the cluster.

The torque, which acts on the magnetic cluster due to
spin-polarization of electric current is a monotonic function of
the distance $r_{0}$ (\ref{Ty}) to linear approximation in the
exchange constant
\begin{equation}
T_{y}\sim \sin \alpha P_{\mathrm{eff}}\widetilde{J}\left(
z_{0}/k_{\mathrm{F} }r_{0}^{2}\right) ^{2},\quad
k_{\mathrm{F}}r_{0}\gg 1,
\end{equation}
where $z_{0}$ is the depth of the cluster below metal surface.

These results may be exploited in future experiments for detecting and
investigating individual magnetic clusters buried below the surface of a
host metal. Specifically, a comparison of the amplitude of the conductance
oscillations for different directions of the cluster magnetic moment allows
determination of the exchange energy $J\mu _{\mathrm{eff}}$ and for a known
value of the exchange integral $J$ to find $\mu _{\mathrm{eff}}.$

\begin{acknowledgments}
One of us (Yu.K) would like to acknowledge useful discussions with B.I.
Belevtsev, A.B. Beznosov, N.F. Kharchenko and D.I. Stepanenko. This research
was supported partly by the program "Nanosystems nanomaterials, and
nanotechnology" of National Academy of Sciences of Ukraine and Fundamental
Research State Fund of Ukraine (project F 25.2/122).
\end{acknowledgments}

\end{document}